\documentclass[twoside]{dis08}
\usepackage[latin1]{inputenc}
\usepackage[dvips]{graphicx,epsfig,color}
\usepackage{wrapfig,rotating}
\usepackage{amssymb,amsmath,array}

\catcode`@=11
\newcount\@tempcntc
\def\@citex[#1]#2{\if@filesw\immediate\write\@auxout{\string\citation{#2}}\fi
  \@tempcnta\z@\@tempcntb\m@ne\def\@citea{}\@cite{\@for\@citeb:=#2\do
    {\@ifundefined
       {b@\@citeb}{\@citeo\@tempcntb\m@ne\@citea\def\@citea{,}{\bf ?}\@warning
       {Citation `\@citeb' on page \thepage \space undefined}}%
    {\setbox\z@\hbox{\global\@tempcntc0\csname b@\@citeb\endcsname\relax}%
     \ifnum\@tempcntc=\z@ \@citeo\@tempcntb\m@ne
       \@citea\def\@citea{,}\hbox{\csname b@\@citeb\endcsname}%
     \else
      \advance\@tempcntb\@ne
      \ifnum\@tempcntb=\@tempcntc
      \else\advance\@tempcntb\m@ne\@citeo
      \@tempcnta\@tempcntc\@tempcntb\@tempcntc\fi\fi}}\@citeo}{#1}}
\def\@citeo{\ifnum\@tempcnta>\@tempcntb\else\@citea\def\@citea{,}%
  \ifnum\@tempcnta=\@tempcntb\the\@tempcnta\else
   {\advance\@tempcnta\@ne\ifnum\@tempcnta=\@tempcntb \else \def\@citea{--}\fi
    \advance\@tempcnta\m@ne\the\@tempcnta\@citea\the\@tempcntb}\fi\fi}
\catcode`@=12

\begin{document}
\title{Status of AKK Fragmentation Functions}

\author{Bernd A. Kniehl
%
%
\vspace{.3cm}\\
%
II. Institut f\"ur Theoretische Physik, Universit\"at Hamburg,\\
Luruper Chaussee 149, 22761 Hamburg, Germany
%
}

\maketitle

\begin{abstract}
We summarize the improvements to the previous AKK extraction of fragmentation
functions for $\pi^\pm$, $K^\pm$, $ p/\bar{p}$, $K_S^0$ and 
$\Lambda/\overline{\Lambda}$ hadrons at next-to-leading order implemented in
the AKK08 sets.
\end{abstract}

\section{Introduction}

In the framework of the QCD-improved parton model, the inclusive production of
single hadrons is described by means of fragmentation functions (FFs)
$D_a^h(x,\mu^2)$.
At leading order, the value of $D_a^h(x,\mu^2)$ corresponds to the probability
for the parton $a$ produced at short distance $1/\mu$ to form a jet that
includes the hadron $h$ carrying the fraction $x$ of the longitudinal momentum
of $a$.
Unfortunately, it is not yet possible to calculate the FFs from first
principles, in particular for hadrons with masses smaller than or comparable
to the asymptotic scale parameter $\Lambda$.
However, given their $x$ dependence at some energy scale $\mu$, the evolution
with $\mu$ may be computed perturbatively in QCD using the timelike
Dokshitzer-Gribov-Lipatov-Altarelli-Parisi (DGLAP) equations \cite{gri}.
This allows us to test QCD quantitatively within one experiment observing
single hadrons at different values of center-of-mass energy $\sqrt s$ (in
the case of $e^+e^-$ annihilation) or transverse momentum $p_T$ (in the case
of scattering).
Moreover, the factorization theorem guarantees that the $D_a^h(x,\mu^2)$
functions are independent of the process in which they have been determined
and represent a universal property of $h$.
This enables us to make quantitative predictions for other types of
experiments as well.

Our previous determinations of light-hadron FFs, yielding the BKK \cite{BKK},
KKP \cite{KKP}, and AKK \cite{AKK} sets, were based entirely on $e^+e^-$
annihilation data in the large-$x$ range, $x>0.1$, where $x=2E_h/\sqrt s$ is
the scaled hadron energy. 
In this presentation, we summarize the improvements from new theoretical
input and additional experimental data implemented in the 2008 update of the
AKK sets for charged pions, charged and neutral kaons, (anti)protons,
and (anti)lambdas (AKK08) \cite{AKK08}.
In Secs.~\ref{sec:theory} and \ref{sec:experiment}, we discuss the most
important improvements due to new theoretical and experimental inputs,
respectively.
In Sec.~\ref{sec:comparison}, we briefly mention how the AKK08 FFs compare
with the AKK ones and those recently presented by Hirai, Kumano, Nagai, and
Sudoh (HKNS) \cite{HKNS} and by de Florian, Sassot and Stratmann (DSS)
\cite{DSS}.
In Sec.~\ref{sec:outlook}, we present an outlook.

\section{Theoretical improvements}
\label{sec:theory}

\begin{figure}[ht]
\begin{center}
\begin{tabular}{cc}
\includegraphics[angle=-90,width=0.48\textwidth]{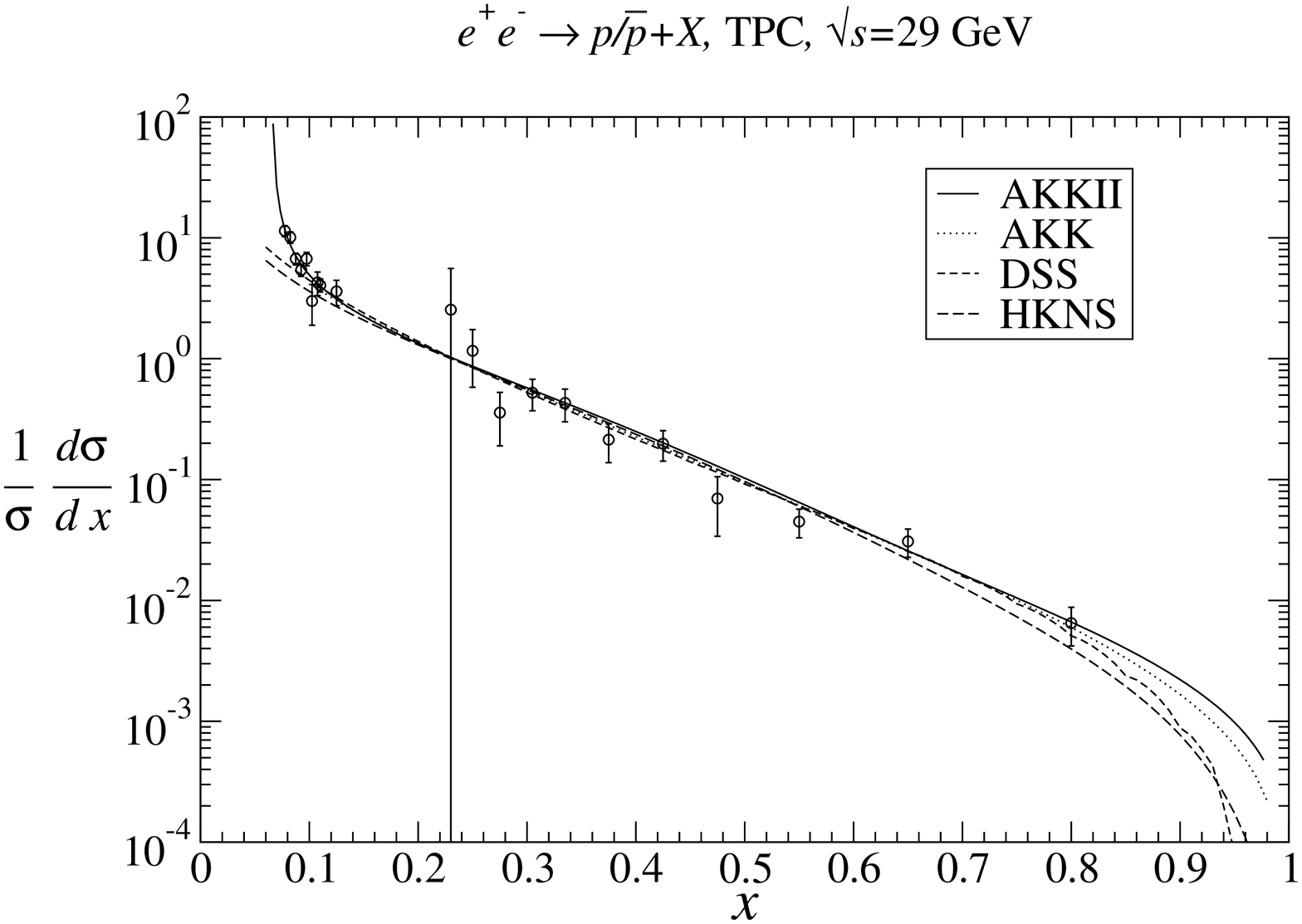}
&
\includegraphics[angle=-90,width=0.48\textwidth]{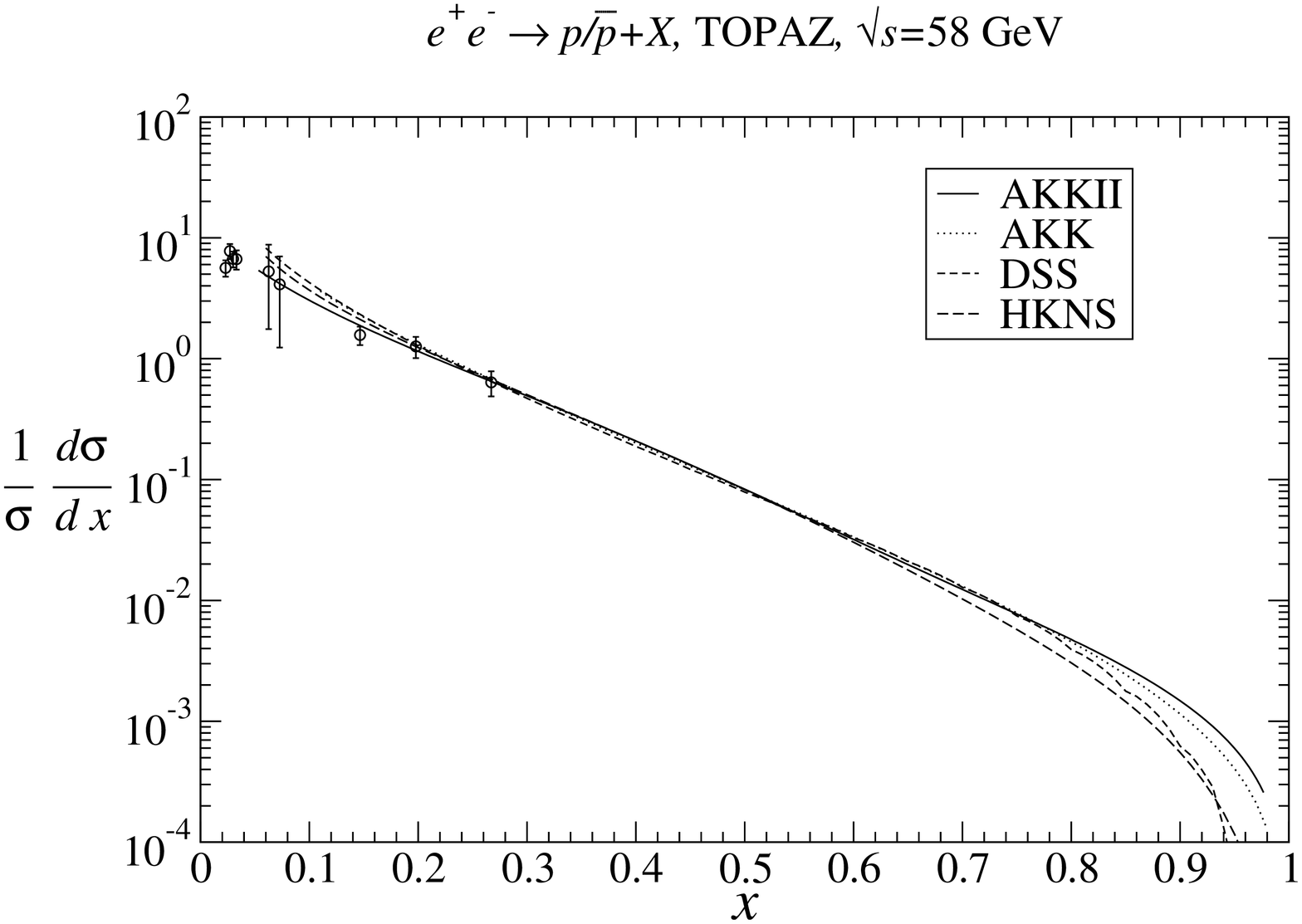}
\end{tabular}
\end{center}
\vspace{-1.cm}
\caption{TPC \cite{TPC} and TOPAZ \cite{TOPAZ} measurements
of $(1/\sigma)(\mathrm{d}\sigma/\mathrm{d}x)(e^+e^-\to p/\overline{p}+X)$
compared with AKK08 \cite{AKK08}, AKK \cite{AKK}, HKNS \cite {HKNS}, and DSS
\cite{DSS} results.}
\label{fig:masses}
\end{figure}

We incorporate hadron mass effects~\cite{Albino:2005gd}, and fit the
hadron mass in the case of the $e^+e^-$ analysis, which also has the
effects of subtracting out other low-$\sqrt{s}$ and small-$x$ effects
beyond the fixed-order approach, such as higher-twist, small-$x$
logarithms, etc.
The fitted hadron masses are presented in Table~\ref{tab:masses}, where they
are compared with the true values.
In the case of the baryons, the fitted masses are
about 1\% above the true values, which is consistent with scenarios in
which the baryons are produced mainly from direct partonic
fragmentation with a small contribution from decays from slightly
heavier resonances.
\begin{wraptable}{l}{0.4\columnwidth}
\centerline{\begin{tabular}{|l|r|r|}
\hline
Particle & Fitted & True \\
\hline
$\pi^\pm$ &  154.6 &  139.6 \\
$K^\pm$ &  337.0 &  493.7 \\
$p/\bar{p}$ &  948.8 &  938.3 \\
$K^0_S$ &  343.0 &  497.6 \\
$\Lambda/\overline{\Lambda}$ & 1127.0 & 1115.7 \\
\hline
\end{tabular}}
\caption{Values of hadron masses (in MeV) resulting from the charge-sign-unidentified 
AKK08 \cite{AKK08} fits of $e^+e^-$ annihilation data and true values.}
\label{tab:masses}
\end{wraptable}
A greater excess is found for the pion mass,
suggesting contributions to the sample from decays of heavier
particles such as $\rho(770)$.
The charged- and neutral-kaon masses are
significantly below their true values.
A possible explanation for
this is that there are significant contributions from complicated
decay channels, such that the direct partonic fragmentation approach
is insufficient.
For this reason we do not impose SU(2) isospin
symmetry of $u$ and $d$ quarks between charged and neutral kaons,
which we do for pions.
In Fig.~\ref{fig:masses}, the TPC \cite{TPC} and TOPAZ \cite{TOPAZ}
measurements of
$(1/\sigma)(\mathrm{d}\sigma/\mathrm{d}x)(e^+e^-\to p/\overline{p}+X)$ are
compared with the calculations using the AKK08, AKK, HKNS, and DSS FFs.
We observe that, owing to the inclusion of hadron-mass effects, the AKK08 set
leads to the best description of both data sets in the low-$x$ range.

\begin{figure}[ht]
\begin{center}
\begin{tabular}{cc}
\includegraphics[angle=-90,width=0.48\textwidth]{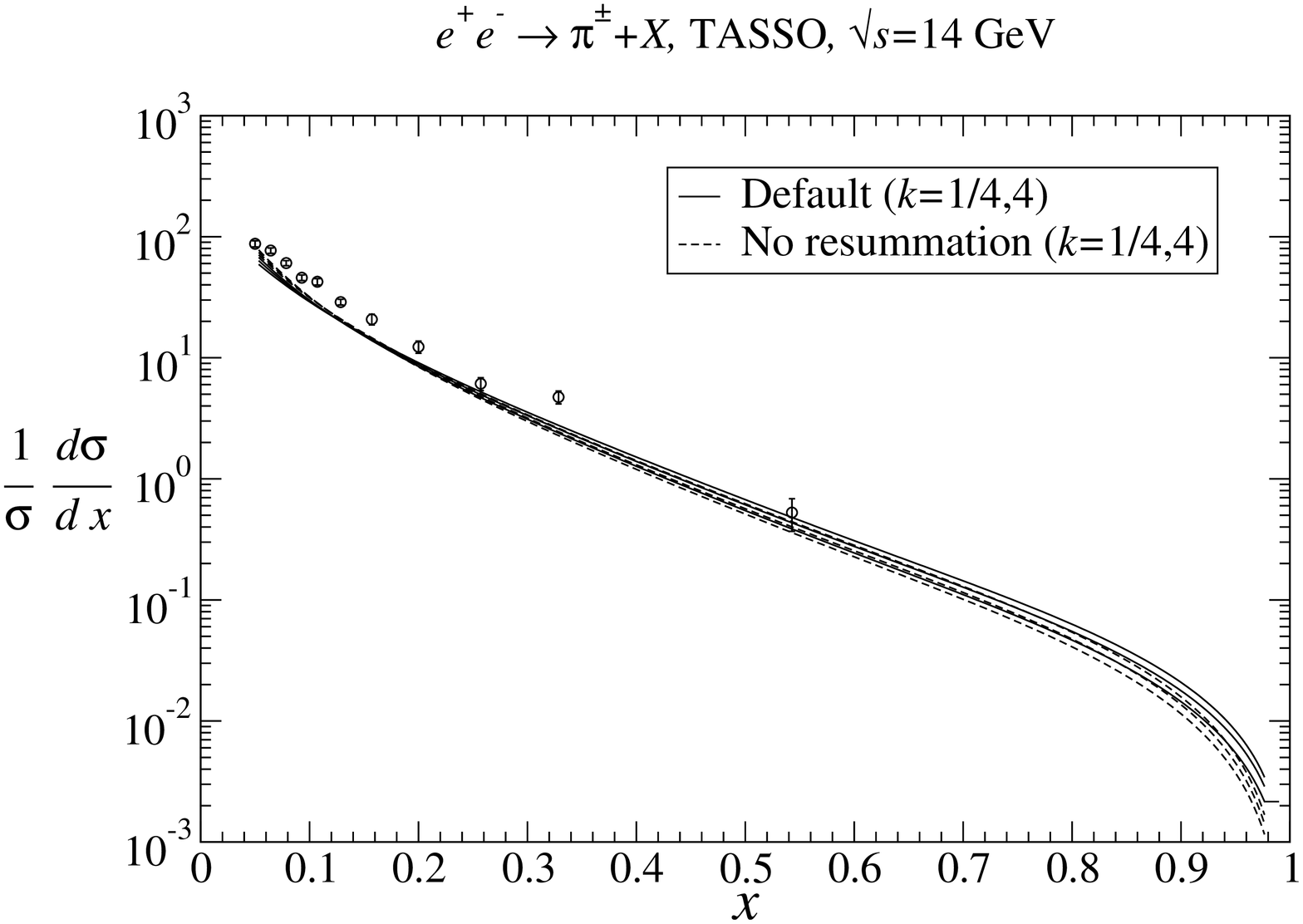}
&
\includegraphics[angle=-90,width=0.48\textwidth]{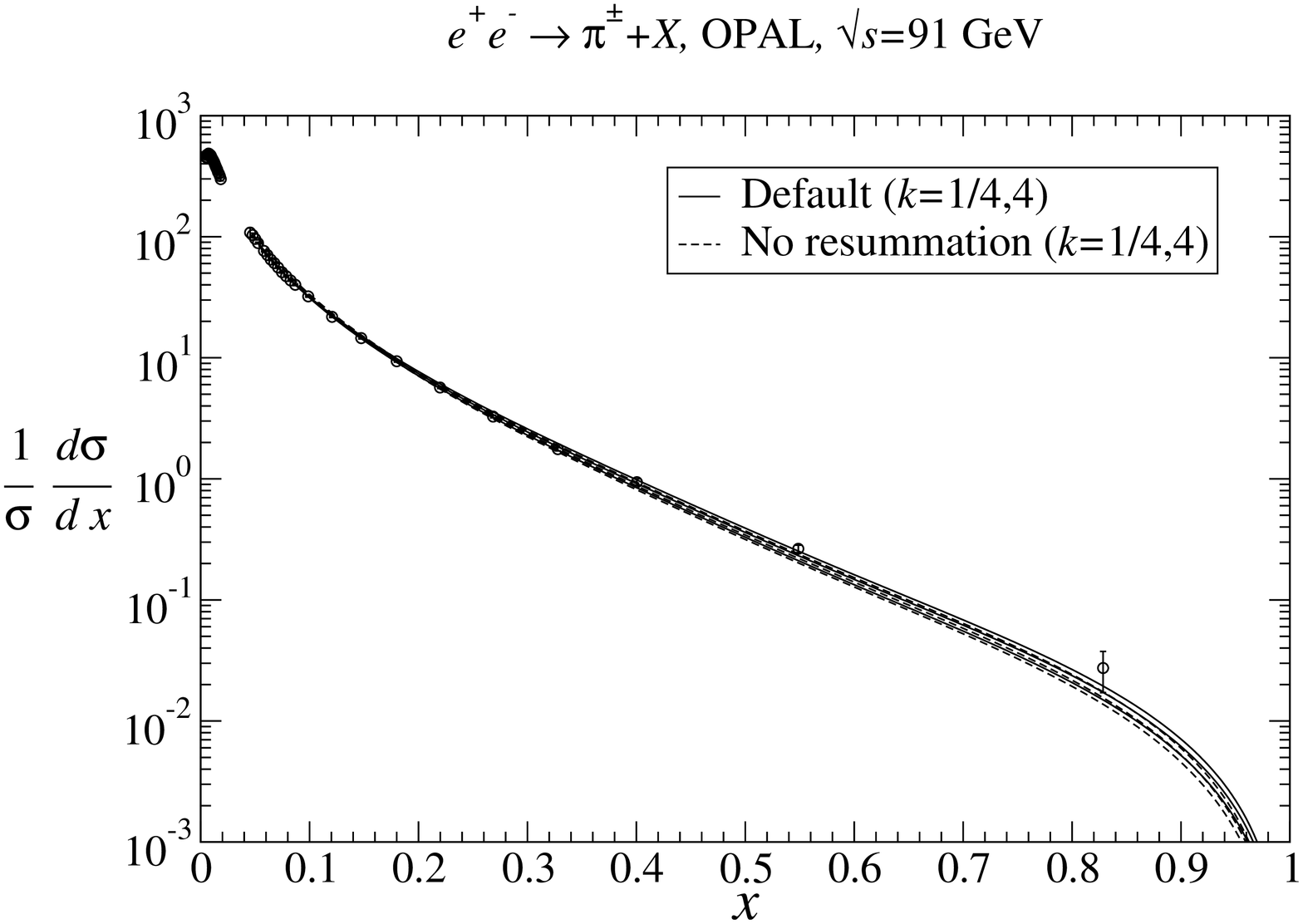}
\end{tabular}
\end{center}
\vspace{-1.cm}
\caption{TASSO \cite{TASSO} and OPAL \cite{OPAL} measurements
of $(1/\sigma)(\mathrm{d}\sigma/\mathrm{d}x)(e^+e^-\to\pi^\pm+X)$
compared with default AKK08 \cite{AKK08} results for scale choice
$k=M_f^2/s=1/4,1,4$ and with the analogous results without large-$x$
resummation.}
\label{fig:largex}
\end{figure}

\begin{wraptable}{l}{0.4\columnwidth}
\centerline{\begin{tabular}{|l|r|r|}
\hline
Particle & Default & Unres.\ \\
\hline
$\pi^\pm$ &  519.7 &  520.8 \\
$K^\pm$ &  417.3 &  488.2 \\
$p/\bar{p}$ &  525.1 &  538.0 \\
$K^0_S$ &  318.6 &  318.8 \\
$\Lambda/\overline{\Lambda}$ &  272.3 &  326.0 \\
\hline
\end{tabular}}
\caption{Minimized $\chi^2$ values in the charge-sign unidentified AKK08
\cite{AKK08} fits and analogous results without large-$x$ resummation.}
\label{tab:largex}
\end{wraptable}
We implement large-$x$ resummation in the
quark coefficient function of $e^+ e^-$ reactions using the results
from Ref.~\cite{Cacciari:2001cw}, since this is a simple improvement
which modifies the cross section over the whole range in $x$ that we
constrain.
Large-$x$ resummation is also implemented in the DGLAP
evolution of the FFs \cite{Albino:2007ns}.
As may be seen from Table~\ref{tab:largex}, this results in a
significant improvement in the fit for charged kaons, (anti)protons
and (anti)lambdas, while $\chi^2$ is essentially unchanged for
charged pions and neutral kaons.
In Fig.~\ref{fig:largex}, the TASSO \cite{TASSO} and OPAL \cite{OPAL}
measurements of $(1/\sigma)(\mathrm{d}\sigma/\mathrm{d}x)(e^+e^-\to\pi^\pm+X)$
are compared with the default AKK08 \cite{AKK08} results for the
choices $k=M_f^2/s=1/4,1,4$ of factorization scale and with the analogous
results without large-$x$ resummation.
We observe that large-$x$ resummation improves the description of the
experimental data in the region of intermediate to large $x$ values.

\section{Additional experimental input}
\label{sec:experiment}

In addition to the constraints of the previous AKK fit, we impose
further constraints on the charge-sign-unidentified FFs from the data
for single inclusive production of identified particles ($p_T \geq 2$~GeV)
from RHIC~\cite{RHIC}
the Tevatron~\cite{Acosta:2005pk}, and
$e^+e^-$ reactions below the $Z$-boson pole and in the range
$0.05<x<0.1$.
While the untagged measurements from $e^+e^-$
reactions provide excellent constraints for the sums of the
charge-sign unidentified FFs for quarks of the same electroweak
charges, they do not constrain the remaining degrees of freedom at
all.
As in the previous AKK fit, these were constrained using quark-tagged data,
while, in the new AKK08 fit, additional constraints are
provided by the data from RHIC.
These data are also much more
sensitive to gluon fragmentation and impose (exclusively) new
constraints on the charge-sign asymmetry FFs.
Normalization errors
were treated as systematic effects, {\it i.e.}\ these errors were
incorporated via a correlation matrix.
Their weights were fitted
analytically and independently of the fit in order to further
ascertain the quality of the fit, and their magnitudes
were typically found to lie in the reasonable range of 0--2.

While the results for the fitted masses suggest that the baryons are
the best candidates for studying direct partonic fragmentation, there
unfortunately exist some inconsistencies between the calculation and
the measurements of the inclusive production of these particles at RHIC:
The description of the STAR data for
$\Lambda/\overline{\Lambda}$ fails, while the contribution from the
initial protons' valence $d$ quarks to the charge-sign asymmetry for
$p/\overline{p}$ from STAR is negative.
Furthermore, while the contributions from the valence and sea-quark
components of the initial protons to the cross section for charge-sign
unidentified particle production at RHIC exhibit the expected behavior,
the contribution from the valence $d$-quark component to the charge-sign
asymmetry in (anti)proton production is negative.
All these issues would be better understood in the
context of an error analysis of the FFs.

\section{Comparisons with other FF determinations}
\label{sec:comparison}

Comparing our FF sets to the recent HKNS \cite{HKNS} and DSS \cite{DSS} ones,
we typically find reasonable agreement for favored FFs, but not for
unfavored ones, and large discrepancies exist at large $x$ in some cases
(see Fig.~\ref{fig:masses}).

\section{Outlook}
\label{sec:outlook}

Future hadron-identified data from BABAR, CLEO, HERA, and RHIC will help
to clarify the open issues mentioned above and significantly reduce the large
uncertainties in much of the FF degrees of freedom.
As for the analysis of high-$Q^2$ electroproduction data from HERA,
hadron species and charge identification would be very useful.

\section*{Acknowledgments}

The author thanks S. Albino and G. Kramer for the collaboration on the
work presented here.
This work was supported in part by the Deutsche Forschungsgemeinschaft
through Grant No.\ KN~365/5--1 and by the Bundesministerium f\"ur Bildung und
Forschung through Grant No.\ 05~HT6GUA.


\begin{footnotesize}

\end{footnotesize}

\end{document}